\documentclass[a4paper,11pt]{article}
\usepackage{graphicx}

\author{M. Cosentino Lagomarsino, F. Capuani\footnote{ FOM Institute
    for Atomic and Molecular Physics (AMOLF), Kruislaan 407, 1098 SJ
    Amsterdam, The Netherlands. e-mail:cosentino-lagomarsino@amolf.nl,
    capuani@amolf.nl.}, \\C.P. Lowe\footnote{Universiteit van
    Amsterdam, Nieuwe Achtergracht 166 NL-1018 WV Amsterdam The
    Netherlands. e-mail: lowe@science.uva.nl.}}

\title{A simulation study of the dynamics of a driven filament in
  an Aristotelian fluid}

\begin{document}
   
\maketitle

{\large 
\begin{center}
    PACS
\end{center}
}

\begin{abstract}
  We describe a method, based on techniques used in molecular
  dynamics, for simulating the inertialess dynamics of an elastic
  filament immersed in a fluid. The model is used to study the
  "one-armed swimmer". That is, a flexible appendage externally
  perturbed at one extremity.  For small amplitude motion our
  simulations confirm theoretical predictions that, for a filament of
  given length and stiffness, there is a driving frequency that is
  optimal for both speed and efficiency. However, we find that to
  calculate absolute values of the swimming speed we need to slightly
  modify existing theoretical approaches.  For the more realistic case
  of large amplitude motion we find that while the basic picture
  remains the same, the dependence of the swimming speed on both
  frequency and amplitude is substantially modified.  For realistic
  amplitudes we show that the one armed swimmer is comparatively
  neither inefficient nor slow.  This begs the question, why are there
  little or no one armed swimmers in nature?
\end{abstract}

\section{Introduction}

For a class of biologically important polymeric materials elasticity
is crucial. Their typical lengths (microns or less) are comparable
with the scale on which rigidity prevents them from collapsing. The
cytoskeletal filaments actin and microtubules \cite{Bray} fall in
this category, as do cilia and flagella. The latter are motile
assemblies of microtubules and other proteins.  Because of their size
and typical velocities, the motion of these filaments is nearly always
in the low Reynolds number regime. This is an inertialess,
Aristotelian, world where the dynamics of a surrounding fluid become
time-reversible. As a notable consequence, it is difficult to generate
any propulsion on this scale~\cite{purcell}.  Nonetheless,
cytoskeletal filaments are involved in cellular and microorganism
motility. Perhaps the most widely known example is that of the
flagellum of a sperm cell, that enables it to swim along the ovaric
tubes. The internal drive of a flagellum, however, is rather
complicated \cite{Sleigh}. It involves many internal degrees of
freedom and active components. On the other hand, modern
micromanipulation techniques, such as optical and magnetic trapping,
open up the possibility of perturbing otherwise passive filaments with
a simplified and controlled drive. This provides a potentially useful
model system for which one may study the fundamentals of motility.

%figura intro
\begin{figure}[htbp]
  \centering
  \includegraphics[angle=-90, scale=.8]{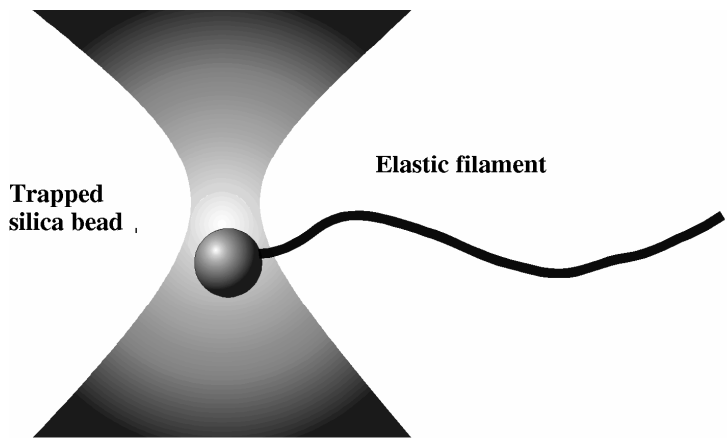}
  \caption{Schematic of a micromanipulation experiment
that allows one to apply a controlled drive to a inert
filament.}
  \label{fig:zero}
\end{figure}

It is this problem we concern ourselves with here. Specifically, we
consider the flexible one-armed swimmer. That is, an elastic filament
that is wiggled at one end. If the filament were rigid, the reversible
motion of the surrounding fluid would ensure that this mechanism
generates no propulsion (the ``scallop theorem'' as Purcell termed
it). However, the flexibility of the arm breaks the time reversal
symmetry for the motion of the assemblage. This makes propulsion, in
principle, possible.  For any microscopic filament the factors that
determine its dynamic behavior are the same. Namely, the equations of
motion will be essentially inertialess. The motion itself will be
determined by a balance between forces driving the filament, friction
forces exerted as the surrounding fluid opposes any motion, and
bending forces that try to restore the (straight) equilibrium state.
For relatively simple model systems, there has recently been
theoretical progress in solving analytically the ``hyperdiffusion''
equation that, in the limit of small amplitude motion, describes the
movement of such a filament.  Wiggins and Goldstein~\cite{WigginsPRL}
considered the motion of a single filament driven at one end by an
external perturbation.  Their analysis emphasized that there are two
very different regimes; one where bending forces dominate and the
filament behaves like a rigid rod, and a second where the viscous
damping of the fluid has the effect of suppressing the propagation of
elastic waves.  For the one armed swimmer, this leads to an optimal
set of parameters that maximize either the swimming speed or swimming
efficiency.  The same analysis gives predictions for the shape of such
a wiggled filament that can be compared with the response observed in
a micro-manipulation experiment~\cite{WigginsBJ}. By comparing
experimental results with theory, structural properties of the
filament were inferred.

Wiggins and Goldstein consider the flexible one armed swimmer in the
limit of small amplitude motion.  With the assumption of inertialess
dynamics, one can derive the following equation \cite{WigginsPRL}
\cite{julicher} \cite{Machin} for the function \( h(s,t) \),
describing the displacement of the filament from the horizontal axis
as a function of time \(t \) and arclength \(s\).

\begin{equation}
  \xi_{\perp} \partial_t h = - k \Delta^2 h 
  \label{eq:wig}  
\end{equation}

Here \(\xi_{\perp}\) determines the viscous force, treated as simply a
transverse viscous drag and \(k\) the stiffness of the filament.  This
``hyperdiffusion'' equation has to be solved subject to appropriate
boundary conditions (corresponding to different forms of external
driving).  Simple active driving mechanisms could be an oscillating
constraint on the end position, or an oscillatory torque applied at
one extremity.  The former can be regarded as the simplest example of
what has been called \emph{elastohydrodynamics} as it involves the
balance of viscous and elastic forces. The latter is a more plausible
biological mechanism as it involves no net external force.  Both
these mechanisms are considered by Wiggins and Goldstein and we also
consider both here.

To summarize the predictions of the theory, for a given amplitude of
driving the remaining parameters can be grouped together to define a
dimensionless ``Sperm number'', \label{secsp}
\begin{equation}
Sp = {\left( \frac{l^4 \omega \xi^{\perp} }{k} \right)}^{\frac{1}{4}}
\label{sp}
\end{equation}
where \( l \) is the length of the filament and \(\omega\) the
wiggling frequency.

This characterizes the relative magnitudes of the viscous and bending
forces.  A low value implies that bending forces dominate, a high
value viscous forces. As a function of the Sperm number, the theory
predicts
\begin{itemize}
\item[-] The swimming speed and efficiency (defined as the amount of
  energy consumed, relative to the amount of energy required to simply
  drag a passive filament through the fluid at the same velocity), go
  to zero as Sp goes to zero. This is the stiff limit where the
  motion is reversible and the scallop theorem applies
\item[-] At a sperm number Sp$ \simeq 4$ there is a maximum in the
  both the swimming speed and efficiency (although not at exactly the
  same value)
\item[-] At high sperm numbers a plateau region where the speed and
  efficiency become independent of Sp, albeit at values lower than
  the peak.
\end{itemize}

In this paper we describe a numerical model that allows us to simulate
such a driven filament.  With the model we can calculate the dynamics,
free from restrictions such as small amplitude motion, and with
greater scope to specify the type of active forces driving the motion
and the boundary conditions applicable for a given physical situation.
With such a model, we can test theoretical predictions and also study
more complex problems where no analytic solution is available.  Here
we do both. Looking at small amplitude motion we compare with the
theory. Moving on to large amplitude motion we establish to what
extent the small amplitude approximation limits the theory.

%The main problem about this experimental system as a model for
%motility is that the force generated is generally too weak to
%be detected by an optical trap. We conclude the paper discussing
%possible ways to get around the obstacle exploiting the information
%given by the model.

\section{Model and Simulation}\label{secmodel}

Our model solves the equations of motion of a discretized elastic
filament immersed in a low Reynolds number fluid.  Any form of
internal and external forcing can be imposed but we restrict ourselves
here to an active force, acting on one extremity, that is periodic in
time. The hydrodynamics is kept to the approximation of slender body
flow \cite{keller}, where the local velocity-force relation is reduced
to a simple expression in terms of friction coefficients that are
shape and position independent.  They do nonetheless reflect the
difference between friction transverse and longitudinal to the
filament. For the problems we are concerned with here the planar
driving forces produce planar motions. The model would apply equally
well were this not to be the case.

\begin{figure}[htbp]
  \centering
  \includegraphics[scale=.3]{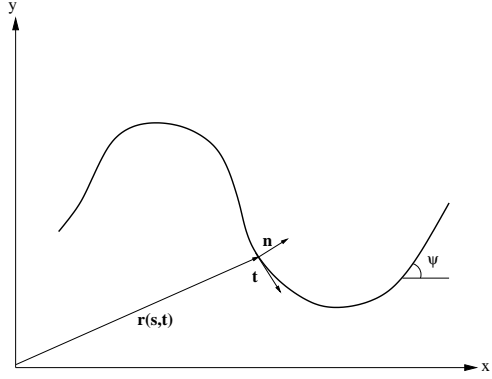}
  \caption{Curve describing the shape of the
  filament. \(\mathbf{n}(s,t) \) and \(\mathbf{t}(s,t)\) are the local
  normal and tangent unit vector respectively. \(\psi(s,t)\) is
  the angle formed with the x axis.}
  \label{fig:shape}
\end{figure}

Considering a continuous description of the filament in space and
time, one can specify this at any given instant \(t\) by a curve \(
\mathbf{r}(s,t) \), giving a point in space for any value of the
arclength parameter \(s\) (figure \ref{fig:shape}).

To describe the dynamics we need the local forces acting on the
filament. The latter are related to the energy of the model
system. Specifically, we have
\begin{itemize}
\item[i.] A bending elasticity, described by the Hamiltonian
  \begin{equation}
    \label{eq:hamiltonian}
    H = \frac{1}{2} k \int_0^L C(s)^2 ds
  \end{equation}
  where \( C(s) = \left(\frac{\partial^2 \mathbf{r}}{\partial
      s^2}\right)^2 \) is the local curvature and \(k \) the
  stiffness.
\item[ii.] A constraint of inextensibility, which can be expressed in terms
  of the tangent vector as 
  \begin{displaymath}
    \left\vert \frac{\partial \mathbf{r}}{\partial s} \right\vert = 1
  \end{displaymath}
  and imposes the condition that the filament is, to a first
  approximation, inextensible.
\item[iii.] An over-damped (negligible mass) equation of motion, which
  can be written as
  \begin{equation}
    \label{eq:motion}
    \partial_t \mathbf{r}(s,t) = - \left( \frac{1}{\xi_{\parallel}}
    \hat{\mathbf{t}}  \hat{\mathbf{t}} + \frac{1}{\xi_{\perp}}
    \hat{\mathbf{n}}  \hat{\mathbf{n}} \right) \frac{\delta H}{\delta
    \mathbf{r}(s)} 
  \end{equation}
  Here, following slender-body theory, the effect of the surrounding
  fluid is taken as a drag force that is proportional and opposite to
  the local velocity. This is anisotropic due to the elongated shape
  of the filament.  This requires the presence of a longitudinal drag
  coefficient \( \xi_{\parallel} \) associated with the projector \(
  \hat{\mathbf{t}} \hat{\mathbf{t}} \) along the tangent vector \(
  \mathbf{t} \), together with a transverse coefficient \( \xi_{\perp}
  \) acting along the normal vector \( \mathbf{n} \).
\end{itemize}

%qui parte analitica
Accordingly, one obtains two equations of motion, one for
the evolution of the filament shape, and the other for the tension
force \(\tau(s,t) \), that enforces locally the inextensibility
constraint.
Expressing the curve shape as the angle \( \psi(s,t) \) that its local
tangent forms with a fixed \( \hat{\mathbf{x}} \) direction, one can
write these equations as (see \cite{julicher}):
\begin{equation}
  \label{eq:psi}
  \partial_t \psi = \frac{1}{\xi_{\perp}} \left( -k \partial_s^4 \psi
  +\tau \partial_s^2 \psi+ \partial_s \psi \partial_s \tau  \right) +
  \frac{1}{\xi_{\parallel}} \partial_s \psi \left( k \partial_s^2 \psi
  \partial_s \psi + \partial_s \tau  \right)
\end{equation}
and
\begin{equation}
  \label{eq:tau}
  \partial_s^2 \tau - \frac{\xi_{\parallel}}{\xi_{\perp}} (\partial_s
  \psi)^2 \tau = -k \partial_s (\partial_s \psi \partial_s^2 \psi) +
  \frac{\xi_{\parallel}}{\xi_{\perp}} \partial_s \psi (-k \partial_s^3 \psi)
\end{equation}

The two nonlinear equations above have then to be solved subject to
appropriate boundary conditions. For example, no external forces and
torques for a free tail.  For the wiggling problems we examine here,
the non-equilibrium drive (oscillating end position or torque) is, in
these terms, simply a time-dependent boundary condition.  Through a
functional expansion about the obvious solution for zero drive \(
\psi_0(s,t) = 0 \), \( \tau_0(s,t) = 0 \),
\begin{equation}
  \begin{array}[h]{cc}
    \psi & = \epsilon \psi_1 + \epsilon^2 \psi_2 + ... \\ 
    \tau & = \epsilon \tau_1 + \epsilon^2 \tau_2 + ...
  \end{array}
  \label{eq:epsilon}
\end{equation}
it is straightforward to obtain, to second order in \( \epsilon \) the
(decoupled) equations
\begin{displaymath}
    \partial_t \psi_1 = - \frac{k}{\xi_{\perp}} \partial_s^4 \psi_1 
\end{displaymath}
for \(\psi_1 \) and
\begin{displaymath}
  \partial_t^2 \tau_2 = -\partial_s(k \partial_s \psi_1 \partial_s^2
  \psi_1 ) -  \frac{\xi_{\parallel}}{\xi_{\perp}} (k \partial_s
  \psi_1 \partial_s^3 \psi_1)
\end{displaymath}
for the tension.  Furthermore, expressing the shape of the filament in
terms of the transverse and longitudinal ``absolute'' displacements
\(u(s,t) \) and \( h(s,t) \) from the direction \( \hat{\mathbf{x}} \)
of the filament's resting position, one gets to equation~\ref{eq:wig}
for the time evolution of \( h \) to second order in \(\epsilon\).

%fine analitica

In the simulations we use a particle model to solve equations
\ref{eq:psi} and \ref{eq:tau} numerically using an approach similar to
molecular dynamics. Time is discretized and the filament is described
as a set of \(n\) point particles rigidly connected by \(n - 1\)
``links''. The interaction between the particles is constructed so as
to reproduce the appropriate collective behavior. For convenience in
implementing the algorithm, we do not simulate the over-damped motion,
given by equation \ref{eq:motion} directly. This would correspond to
the zero mass case. Rather, we solve the damped Newton equation for an
object with ``small'' total mass \(m\).  By making the mass small
enough we can reproduce the required inertia-less mass independent
behavior~\cite{ChrisSp}.

\begin{figure}[htbp]
  \centering
  \includegraphics[scale=.4]{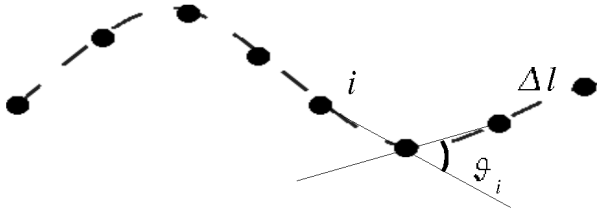}
  \caption{Discretization of the filament}
  \label{fig:discr}
\end{figure}

The bending forces acting on the individual particles are defined
as follows.  If we consider three consecutive discretization points,
their positions will lie on one unique circle of radius, \(R_i\),

\begin{displaymath}
  C_i^2 = {\left( \frac{1}{R_i} \right)}^2
  = \frac{ 2 }{{\Delta l}^2}\left( 1 - \cos( \theta_i ) \right)
\end{displaymath}
where \(\Delta l = L/(n-1)\) is the link length and \(\theta_i\) the angle
between two links at the position of bead \(i\).  We introduce a
bending potential \(U_i\) of the form 
\begin{displaymath}
  U_i = A \left(1 - \cos (\theta_i) \right)
\end{displaymath}
so that the total bending energy will be 
\begin{displaymath}
  H = \sum_{i=2}^{n} U_i = A \sum_{i=2}^{n} \left(1- \cos ( \theta_i ) 
      \right) 
\end{displaymath}
which we can compare with a discretization of the integral in 
\ref{eq:hamiltonian} 
\begin{displaymath}
  H = \frac{k}{2} \Delta l
  \sum_{i=2}^{n} {\left( C_i^2 \right)} 
\end{displaymath}
This leads to the identification \(A = \frac{k}{\Delta l} \).  A more
sophisticated approach~\cite{Vincent}, where the problem is mapped
onto the worm-like chain model of Kratky and Porod~\cite{Porod}, leads
to a slightly different expression,
\begin{math}
A = \frac{n-1/2}{L}k
\end{math}.
The two expressions are equivalent in the limit $n \rightarrow
\infty$, where they reproduce the bending energy of the continuous
filament, but they differ for the finite number of beads used in the
model. The latter leads to faster convergence in the results as the
number of discretization point particles is increased. We therefore
chose to adopt it.

The inextensibility constraint is implemented by introducing equal and
opposite forces along the links between particles. The magnitude of
the forces is computed by imposing a fixed distance \( \Delta l \)
between consecutive beads at each time step. This is a straightforward
matter from the computational point of view, as it involves only the
inversion of a tridiagonal matrix \cite{constraints}.

The viscous drag forces acting on the particles of the model filament
are taken as \( F_{ij} = - \delta_{ij} (\xi_{\parallel} \hat{\bf t}
\hat{\bf t} + \xi_{\perp} \hat{\bf n} \hat{\bf n}) \mathbf{v} \),
where \( \hat{\bf t} \) and \( \hat{\bf n} \) are respectively unit
vectors parallel and normal to the filament, \( \xi_{\parallel},
\xi_{\perp} \) are the longitudinal and transverse friction
coefficients, and \( \mathbf{v} \) is the local velocity. This means
that hydrodynamics is approximated as a local effect on the filament
so the hydrodynamic interaction between different points along the
curve does not vary. The global shape of the curve enters only through
the anisotropy of the viscous drag coefficients acting on individual
points. The ratio of the two coefficients depends on the geometric
details of the filament analyzed. For cilia, flagella, or cytoskeletal
filaments, its value is typically taken between \(1.4\) and
\(2\)~\cite{gueron1}. We chose to adopt an arbitrary \(1.5\) in most
of our simulations, but we also explored different values, including
the cases where the two drags are equal or their ratio is lower than
one.  The time evolution is evaluated in a molecular dynamics-like
fashion, with the only slight subtlety that the Verlet algorithm has
to be modified to allow for the velocity dependent anisotropic viscous
force \cite{ChrisSp}.  Finally, the active drive at the head is simply
implemented as a constraint on the first or first two particles. That
is, \( y_1 = h_0 \cos(\omega t) \) for the oscillating constraint or a
periodic torque, \( T_x = B \sin \left( \omega t \right) \) realized
as a couple of forces applied to the first two beads. Here \( \omega
\) is the driving frequency.

\section{Results for small deviations}\label{secresults}

\subsection{Wave Patterns}

Using the ''Sperm Number'' Sp defined in section \ref{secsp}, we can
characterize the relative magnitude of the viscous and bending forces.
To recapitulate, a low value of Sp indicates that bending forces
dominate, whereas for low values the dominant forces are viscous.  One
reason for defining this number comes from the solution of equation
\ref{eq:wig} \cite{WigginsPRL} \cite{ChrisSp}. In fact, Sp can be
interpreted as a rescaled filament length, where the rescaling factor
is a characteristic length \( l_* = \left( \frac{k}{\omega
    \xi_{\perp}} \right)^{1/4} \) that can be used to
non-dimensionalize the equation.  Both for the oscillating constraint
and oscillating torque we recover the fact that the dynamic response,
for a fixed driving amplitude, is solely dependent on Sp.

\begin{figure}[!phtb]
  \centering
  \includegraphics[scale=.29]{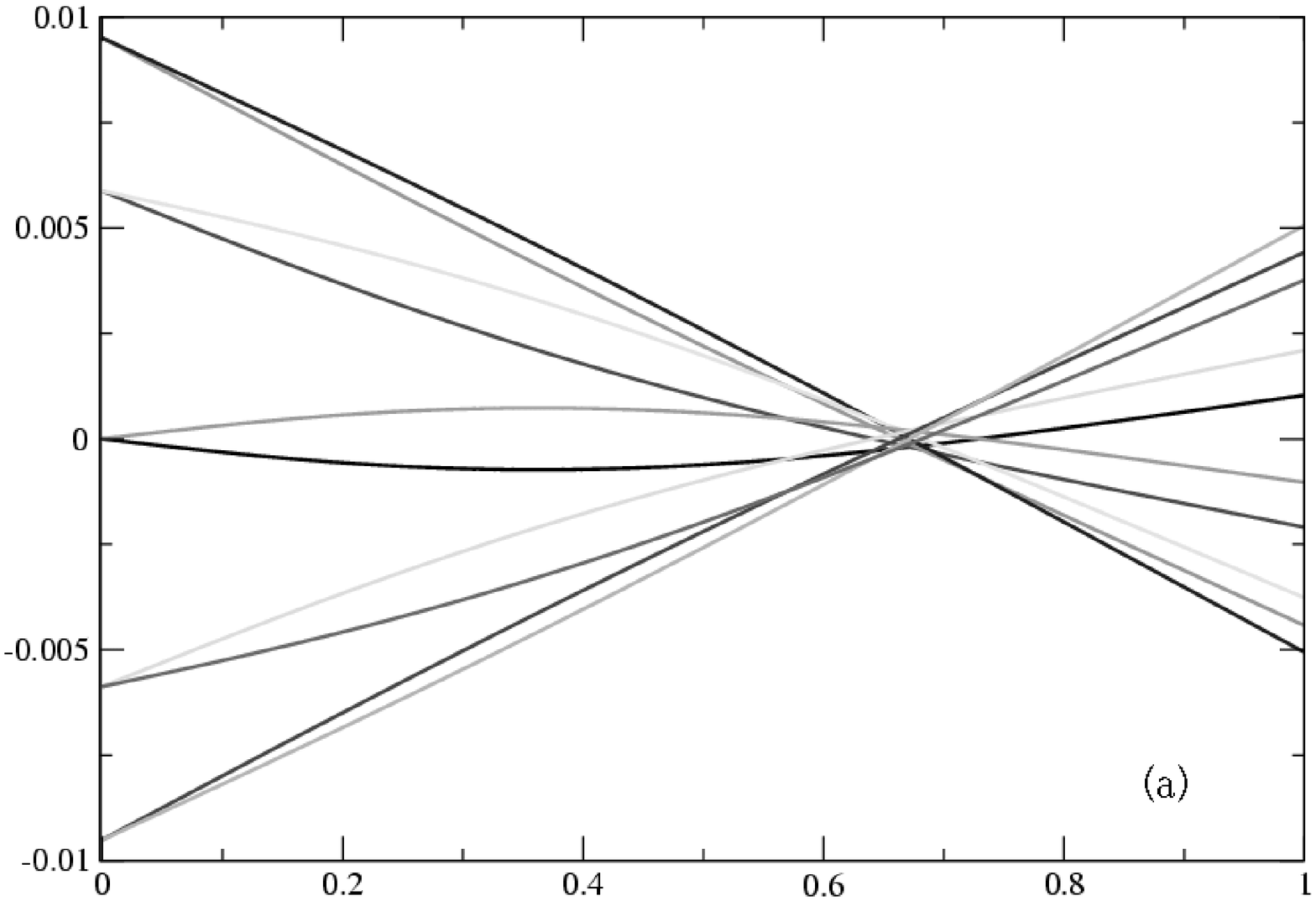}
  \includegraphics[scale=.29]{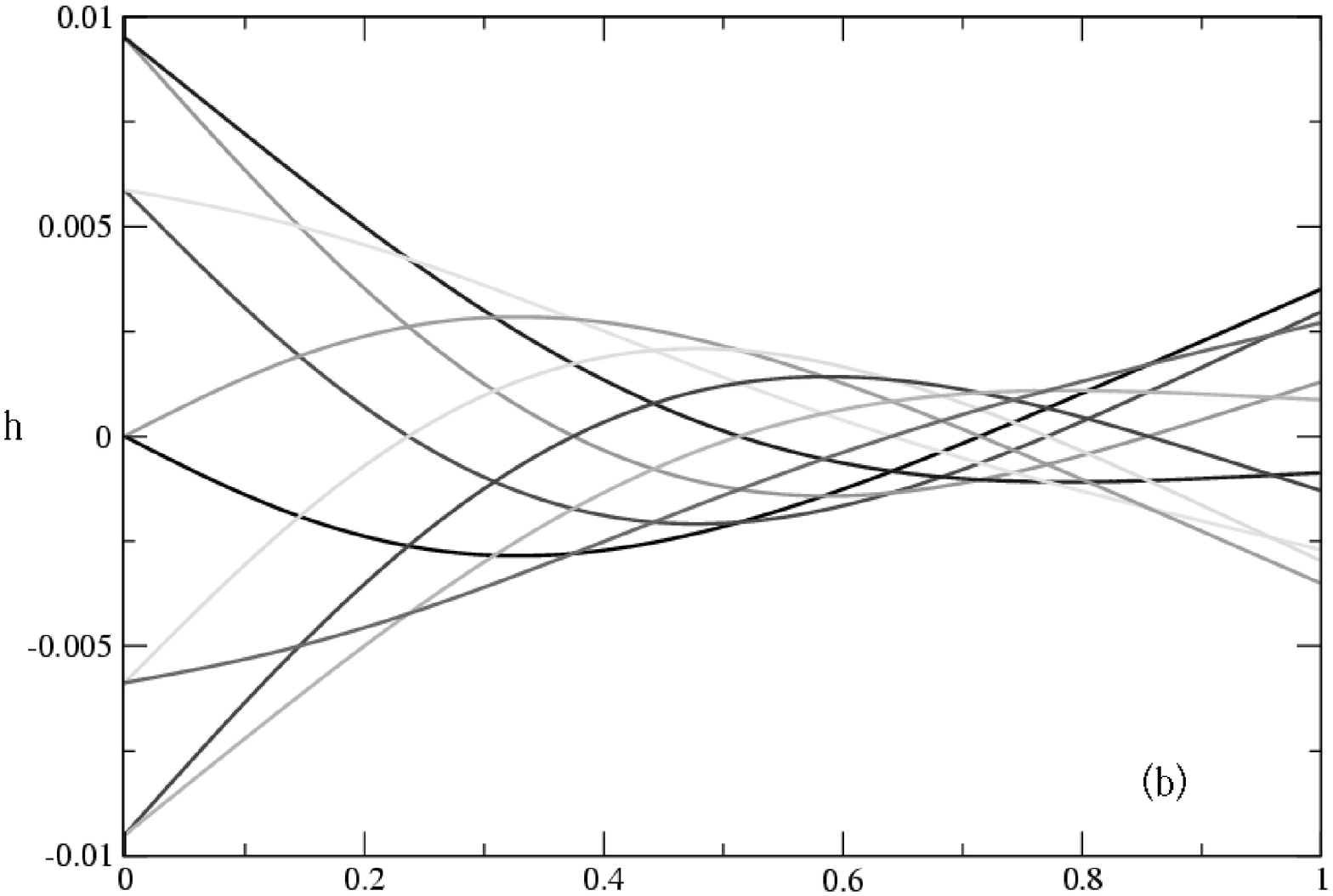}
  \includegraphics[scale=.29]{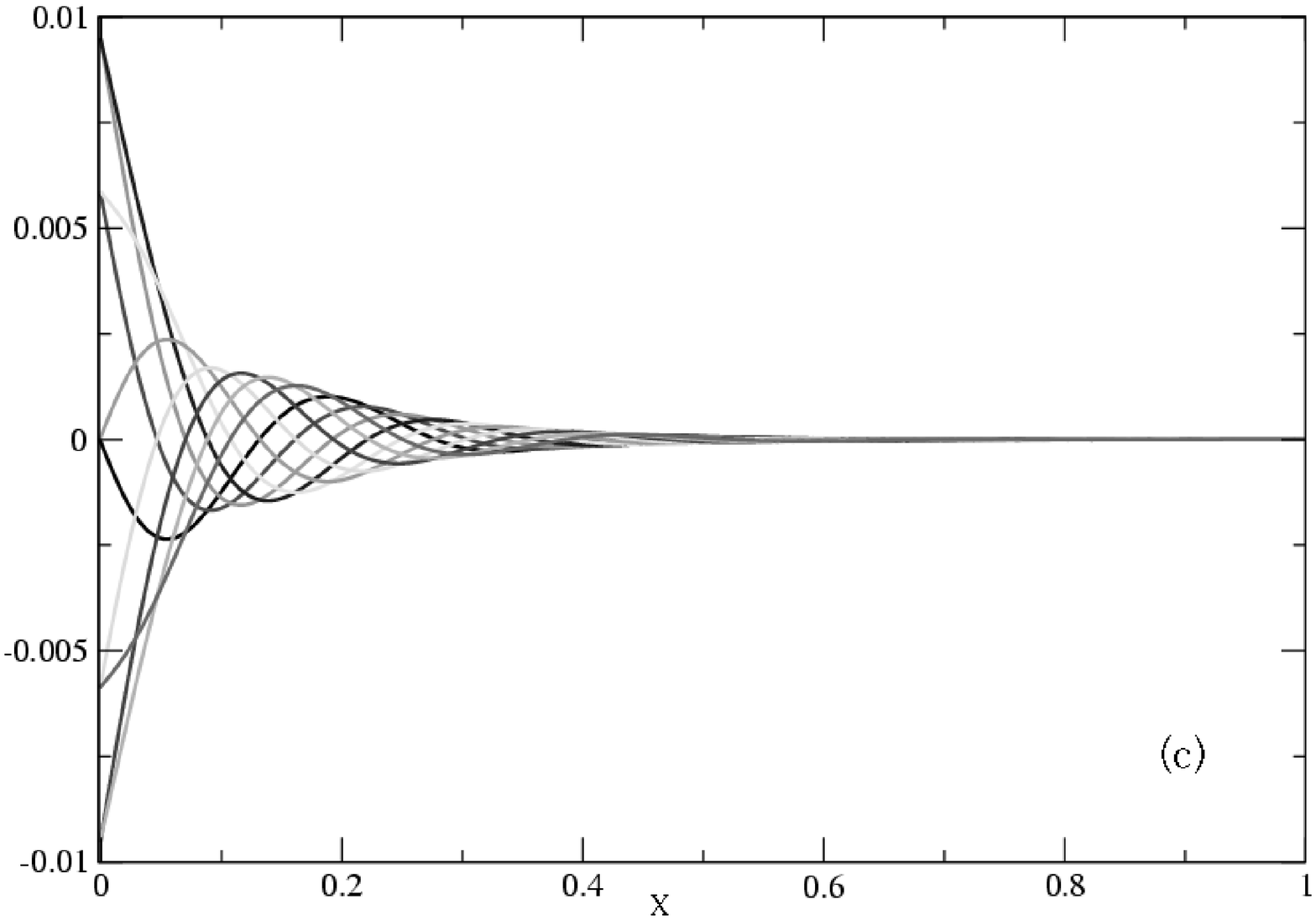}
  \caption{Waveforms of the filament oscillations for different values
    of Sp. (a) Low Sperm number (Sp\(= 2.46\)). The motion consists of
    pivoting oscillations about a fixed point. (b) Intermediate Sperm
    number (Sp\(= 4.29\)). A damped wave propagates along the filament
    making the movement non-reciprocal in time. (c) High Sperm Number
    (Sp\(= 22.61\)).  The propagating wave is damped within a length
    that becomes smaller with increasing Sp.}
  \label{fig:2}
\end{figure}

In figure \ref{fig:2} we have plotted the wave patterns for the
filament at different values of Sp.  These results were obtained using
the oscillating constraint.  That is, the transverse position at the
wiggled end is forced to be sinusoidal in time.  The amplitude of the
motion is small, the maximum displacement being \(1\%\) of the
filament length.  The pictures can be interpreted as ``stroboscopic
snapshots'' of the filament's motion.  For small Sp, bending forces
dominate and the stiff filament pivots around a fixed point. This
motion is virtually symmetric with respect to time inversions
(``reciprocal'').  As Sp increases a (damped) wave travels along the
filament and time reciprocity is broken.  For increasing values of Sp,
viscous forces overcome elastic forces and the characteristic length
scale of damping of the traveling wave becomes smaller. This requires
that the spacing between the beads in our discrete model must also be
reduced to give a fixed degree of accuracy.  The number of beads in
the model (or equivalently the inverse bead spacing) were thus
increased with increasing Sp to ensure that the results are within a
percent of the true, continuum, values.  The oscillating torque gives
qualitatively similar results.

\vspace{.5cm}

All these results are in agreement with the analytical findings of
Wiggins and Goldstein \cite{WigginsPRL} in the small amplitude
approximation. The agreement is also quantitative.

\subsection{Swimming}

From the simulation we are also able to compute the velocity and
efficiency of the movement generated transverse to the wiggling
direction (due to the propulsive force generated by the presence of
the active force) as a function of Sp.  We define swimming of the
immersed object as the generation of motion, through modifications of
shape, in the direction along which no {\em external} force acts.
Both the speed and efficiency, as Wiggins and Goldstein predict,
display an optimum value at intermediate (but different) values of Sp.
Subsequently they reach a plateau as viscous forces begin to dominate
(Sp increases).

\begin{figure}[htbp]
  \centering
  \includegraphics[scale=.4]{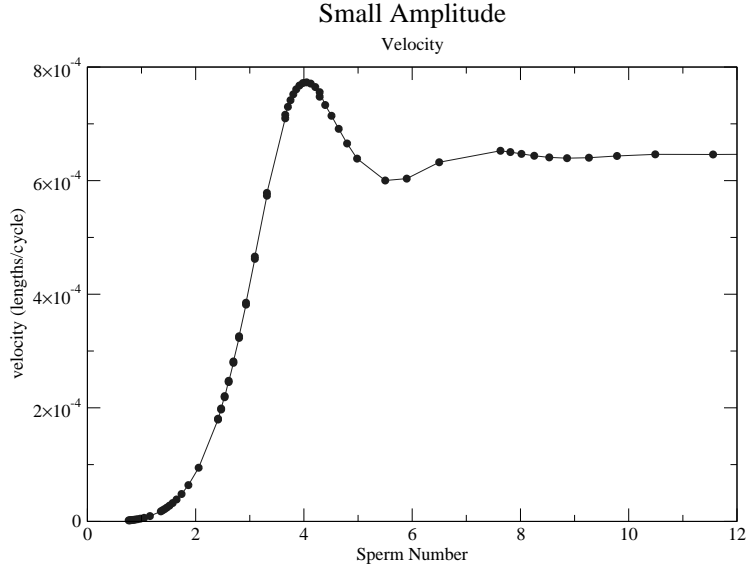}
  \caption{Propulsive velocity as a function of Sperm number for small
    amplitude oscillation (\( h_0 = 0.01 l \)) of the constraint at
    the ''head'' end of the filament. The function goes as the fourth
    power of Sp for small values of this number, reaches an
    optimum, then relaxes to a plateau. }
  \label{fig:3}
\end{figure}

According to the ``scallop theorem'' of low Reynolds number
hydrodynamics, reciprocal (time reversion invariant) motion generates
no swimming \cite{purcell}. This is a consequence of the
time-reversibility of Stokes flow and sets an important condition for
the ability of microorganisms to swim.  In our case, this implies that
we expect no swimming as Sp approaches zero and the motion approaches
reciprocity.  This is confirmed by the result in figure \ref{fig:3}.
The optimum of the velocity is thus the result of a trade-off between
non-reciprocity of the motion and damping of the traveling wave.

\begin{figure}[htb]
  \centering
  \includegraphics[scale=.5]{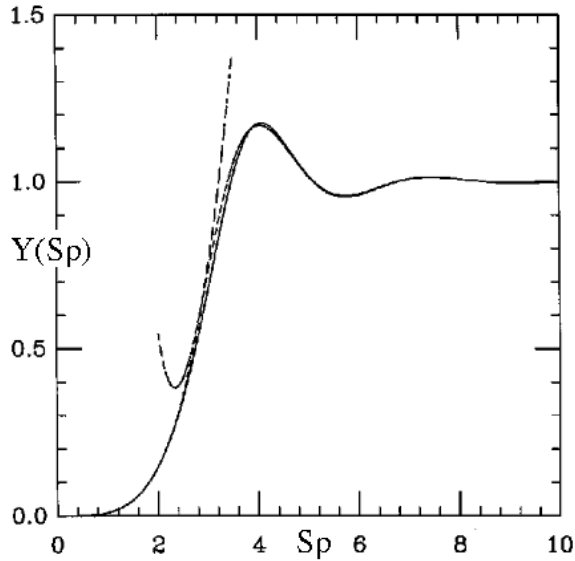}
  \caption{The function Y(Sp) from Wiggins \emph{et al.}.}
  \label{fig:Y(sp)}
\end{figure}

At this point we should also be able to compare our results
quantitatively with those obtained analytically using the
approximation of small deviations.  However, in this respect the
theoretical analysis is somewhat misleading.  Computing the time
average of the force, as in \cite{WigginsBJ} and \cite{WigginsPRL},
yields the expression
\begin{displaymath}
\label{wigspeed}
  \bar{F} = \frac{h_0^2 \xi_{\perp} \omega}{4 \sqrt{2}} \textrm{Y(Sp)}
\end{displaymath}
where Y(Sp) is a scaling function that can be computed exactly (figure
\ref{fig:Y(sp)}).  This expression depends only on the transverse
friction coefficient and does not reduce to zero when \(\xi_{\perp} =
\xi_{\parallel} \).  As such, it is impossible to relate this to the
swimming speed. This follows from the fact that if the condition
\(\xi_{\perp} = \xi_{\parallel} \) is satisfied there can be no
swimming. It is easy to show this must be the case (basically as a
consequence of Newton's third law).  The main reason is that, if one
considers one particle (i.e. a short piece of filament), the effective
viscous drag that it experiences at any moment in time is decoupled
from the local configuration of the filament if there is no anisotropy
in the friction coefficients.  Averaged over one cycle, this always
leads, effectively, to reciprocal motion.  All the forces sum to zero
so there can be no displacement. This is shown more formally in the
Appendix.  Furthermore, our simulations do indeed yield no average
velocity if the two friction coefficients are equal (we use this to
check that there is no ``numerical'' swimming, due to the accumulated
errors in the simulation).  Thus the result given in
equation~\ref{wigspeed}, whilst analytically exact, is misleading
(probably due to subtleties in the formalism of the over-damped
equation of motion).  To correct for this anomaly we used the theory
and computed instead, following the procedure outlined in
\cite{julicher}, the time average of the swimming velocity given the
analytical solution for the shape~\cite{WigginsBJ}.  This yields (see
Appendix)
\begin{equation}
  \label{eq:velocity}
  v = h_0^2 \left( 1 - \frac{ \xi_{\perp} }{ \xi_{\parallel}}\right)
    \frac{\omega}{4 \sqrt{2} L} \textrm{Y(Sp)}
\end{equation}
where Y(Sp) is again the scaling function specified by Wiggins and
Goldstein in computing the average force (figure \ref{fig:Y(sp)}).
Note that this expression (equation~\ref{eq:velocity}) predicts no
swimming when
\begin{itemize}
\item Sp \(=0\) and the motion is reciprocal in time (see fig. 2)
\item When the two drag coefficients \(\xi_{\perp} \) and
  \(\xi_{\parallel} \) are equal.
\end{itemize}
consistent with both the scallop theorem and Newton's third Law.
It also predicts a change in the swimming direction if the friction
coefficients are interchanged.  Curiously, this reversal of direction
has a biological analogue in the organism \emph{Ochromonas} which has
a flagellum decorated by lateral projections (mastigonemes) and swims
in the same direction as that of the propagating wave. The body
follows the flagellum, instead of preceding it as in sperm cells
(~\cite{Bray}, p.11).

Comparing the modified analytical expression for the swimming speed
with the simulations, the essential features predicted are obviously
present. Both approach zero as \( \textrm{Sp}^4 \) for small Sperm
numbers, but with increasing Sperm number display a maximum. In fact,
a careful analysis shows that the agreement, in the small amplitude
limit, is exact.  The presence of a plateau at high Sp is hard to
interpret, in the sense that it predicts velocities for even the
``infinitely floppy'' filament, where the wave pattern is completely
damped in an infinitely small region close to the driven extremity.
However, in our simulations we see the velocity dropping only when the
size of this damping region is comparable to the distance between two
subsequent discretization points, so we have to confirm the analytical
result and explain this oddity, as we will see, as a feature of the
small deviation approximation.

\section{Large Angular Deviations}

\begin{figure}[htb]
  \centering
  \includegraphics[scale=.38]{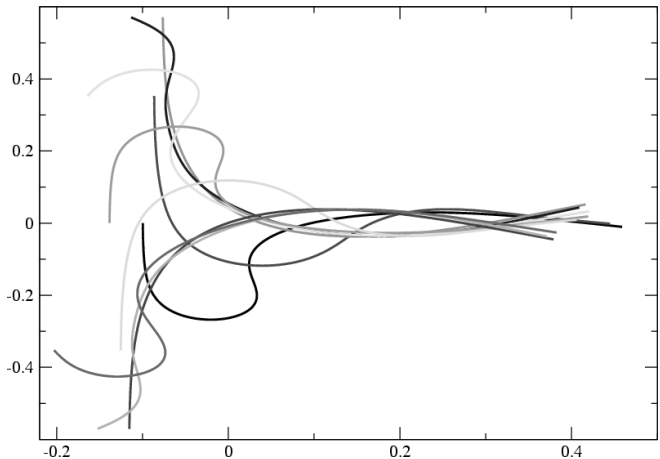}
  \caption{Wave patterns for the case of an oscillating constraint
  with amplitude \( h_0 = 0.6 \  l \) at \(\textrm{Sp}= 2.8\). }
  \label{fig:largewave}
\end{figure}

Our simulation contains the full nonlinear model for the dynamics of
the filament, its only limitation being the discretization of space
and time. Therefore, it is interesting to use it to investigate the
limitations of the analytical model when the motion involves shapes
that deviate significantly from straight. This is also closer to a
real experimental (or biological) situation. The shapes we found often
cannot be described by a function, as the displacement from the
horizontal axis is not single-valued. This can be observed in figure
\ref{fig:largewave}, where we show an example of wave pattern for the
case of oscillating large amplitude constraint.  In this case, the
maximum transverse displacement is $60 \%$ of the tail length.
Looking at this figure, it is obvious that the behavior predicted by
equation \ref{eq:wig} will be substantially modified.

%\vspace{1.5cm}

\begin{figure}[htb]
  \centering \includegraphics[scale=.5]{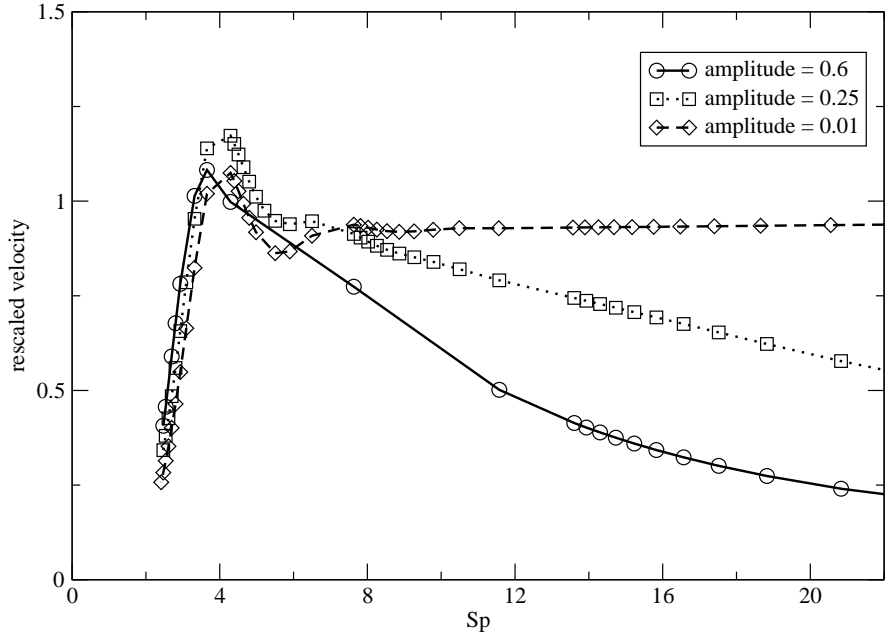}
  \caption{Velocity versus sperm number for different values of the
  amplitude \(h_o\) in the case of oscillating constraint. The
  velocities have been rescaled to show the drop of the plateau.}
  \label{fig:largampli}
\end{figure}

The first notable area of disagreement is at high values of the Sperm
number ($\textrm{Sp} >> 1$) where we no longer find a plateau but a
slow and steady drop in both speed and efficiency (figure
\ref{fig:largampli}).  This effect is clearly a consequence of the
non-negligible amplitude of the motion because for smaller amplitudes
a plateau is indeed reached. This is a limitation of the theory, one
respect in which large amplitude motions differ from the small
amplitude limit.  Further the results for a dimensionless amplitude of
\(0.25\) display a transient plateau that subsequently decays to zero.
This implies that for any finite amplitude the dimensionless swimming
speed always goes zero for large enough Sp. The smaller the amplitude
the longer the plataeu persists, but only for negligible amplitude, is
it the asymptotic behaviour.  It should be noted that figures
\ref{fig:3} and \ref{fig:largampli} should be interpreted with care.
The swimming velocity is plotted in units of the fraction of the
length per cylce.  To obtain absolute swimming speeds, for a tail of
given length and stiffness, we would need to multiply this
dimensioless swimming speed by the frequency.  The frequency itself is
proportional to Sp\(^4\) so a plateau in these plots still implies a
swimming speed increasing proportionally with \(\omega\).  The drop
from the plateau means that the actual swimming speed will increase
with frequency, but a a slower rate. Thus, in practice the one armed
swimmer can go as fast as he or she likes by wiggling fast enough.

\begin{figure}[htb]
  \centering
  \includegraphics[scale=.5]{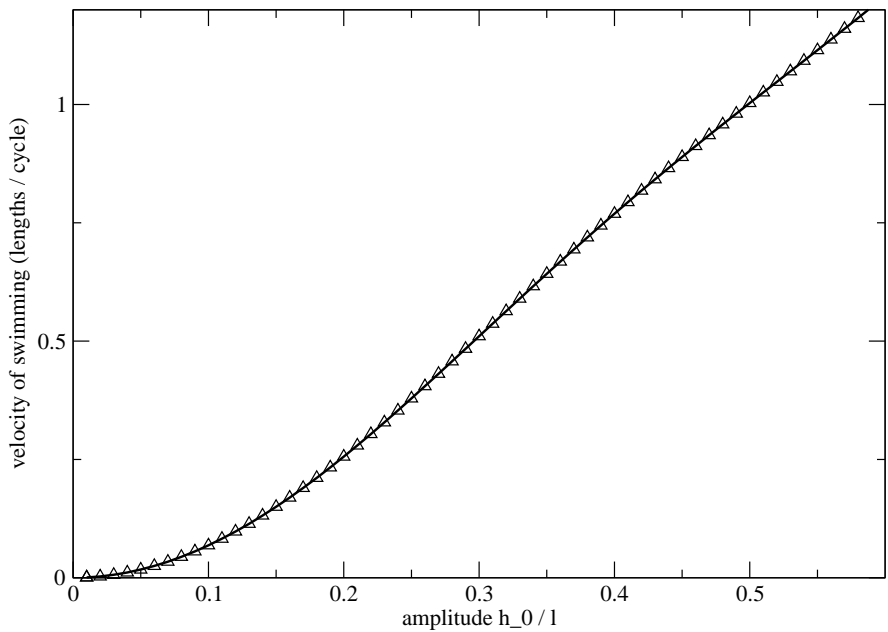}
  \caption{Velocity versus amplitude for the case of an oscillating
  constraint close to the optimum for the speed (Sp = 4). }
  \label{fig:ampli}
\end{figure}

Secondly, we find that the dependence of the optimum swimming speed,
equation \ref{eq:velocity} predicts as the square of the amplitude of
the oscillating constraint, becomes linear for higher amplitude
oscillations (figure \ref{fig:ampli}). 
Thus far, we have not been able to show why this is the case,
but we believe it is related to the following. For small amplitude
motion the elastic wave simply propagates along an essentially straight 
filament. As the amplitude increases, this is no longer true because
the filament itself is significantly bent and, so far as the
damping is concerned, it is the distance along the filament that
is relevant. This is no longer the same quantity as the absolute distance.
This seems to lead to an increase in the effective length of the filament.

From these results it is clear that the amplitude of the drive for
which the small deviation approximation breaks down depends on the
value of Sp, being greater for smaller Sperm Numbers. At the optimal
value for the speed, Sp\(=4\), the approximation holds for maximum
transverse displacements of up to \(20 \%\) of the tail length, which
is well beyond the point one would expect the assumptions to be valid.
However, for a realistic experiment with actin or a microtubule, the
values of Sp are much higher than \(4\), and the value for the
theshold is much lower. For example, for an actin filament of \( 50
\mu m\), driven at 1 cycle/second at an amplitude of \(25\%\) of its
length, we estimated a speed of about \(1,7 \mu m/sec\) with the small
deviation model, whereas our simulation predicts a reduction of this
value by a factor \(1/4\).

\begin{figure}[htb]
  \centering
  \includegraphics[scale=.5]{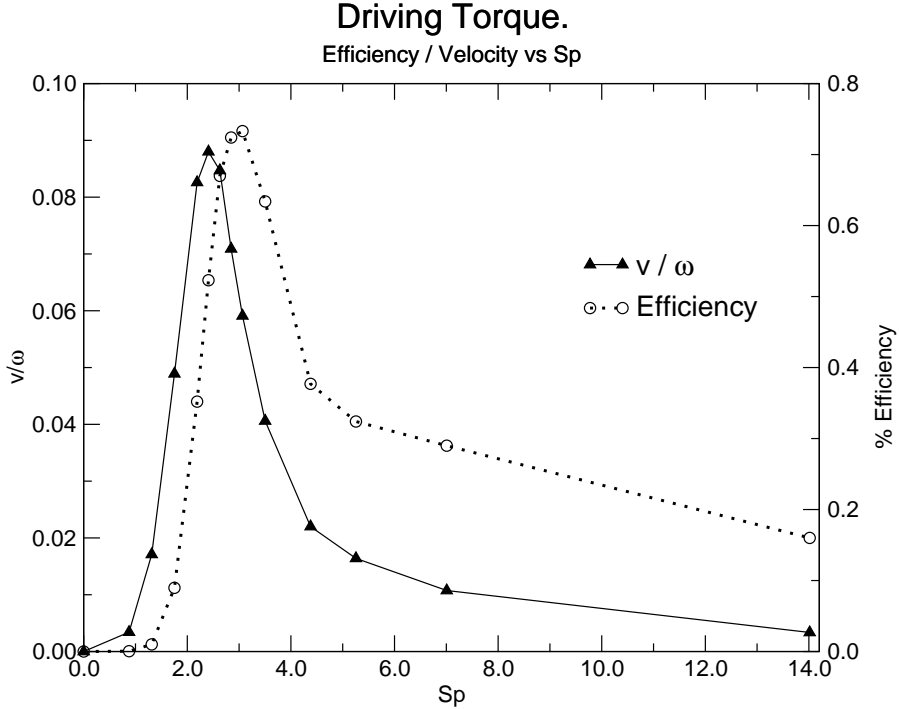}
  \caption{Velocity / efficiency versus Sp for the case of driving
  torque.}
  \label{fig:torque}
\end{figure}

For external driving in the form of a torque applied at one end, we
have only considered large amplitude motion. Specifically, the
pre-factor $B$ was adjusted to produce a maximum angle at the driven
end of $60^o$.  This clearly violates the small angle approximation of
Wiggins and Goldstein~\cite{WigginsPRL} but is more consistent with
the head deflections found in practice for swimming organisms.  In
figure \ref{fig:torque} we plot the efficiency and the mean velocity
as functions of Sp. Once again, the two curves agree qualitatively
with those found analytically by Wiggins and Goldstein in that there
is a peak speed and efficiency. The values are at slightly different
values of Sp and, because the small angle approximation is violated,
not quantitatively predicted by equation~\ref{eq:velocity}.

We should add here a few comments. Notably, the peak efficiency of
less than $ 1 \%$ seems very low. However, this depends strongly on
the amplitude of the motion. If we go above the $60^o$ limit we have
imposed here for the torque, or to driving amplitudes of greater than
half the length of the filament, it is possible to reach values of $2
\%$ before the motion becomes unstable. This is similar to the
efficiency typical for both the helical screw mechanism used by
bacteria and the sperm motion~\cite{ChrisSp}. Thus, the one-armed
swimmer operating at peak efficiency is a plausible and not especially
inefficient entity. Note also that the efficiency (which is
dimensionless), as well as the swimming speed, decays to zero rather
than reaching a plateau value. This means that in absolute terms the
one-armed swimmer can carry on with increasing speed by increasing its
wiggling frequency but only at the price of decreasing efficiency.

\section{Conclusions}
We have described a simulation method that can be used to study the
motion of driven elastic filaments in a low Reynolds number flow.
Here, the hydrodynamic friction is treated quite simply, consistent
with comparing with analytically tractable theories.  A more complete
calculation of the hydrodynamic effects would no doubt be instructive.
In particular, the friction coefficients are not, as we assume here,
independent of distance along the filament.  At the expense of a
little more computational complexity, such effects could be
incorporated into our model in a straightforward manner.
%MARCO: A REF?
We showed that, within this approximation, the picture suggested by
Wiggins and Goldstein for the linear regime of small angular
deviations from the straight position is essentially correct. Our
results for the motion of the filament, show good agreement with their
analytical calculations.  There is an optimal balance between bending
forces and viscous forces that leads to a maximum propulsive speed and
efficiency.  However, in one quantitative respect our results suggest
that their analysis is limited. We could not relate their expression
for the average force exerted by the filament to the swimming speed.
Instead, we used their model to compute an expression for the average
swimming velocity that is physically more plausible and agrees with
the simulation results.

For large amplitude motion, we found that the dependence of the
swimming speed on both sperm number and amplitude was significantly
modified relative to the small amplitude case.  Further, we postulate
that this is due to the fact that in a highly distorted filament the
wave travels along a notably different path than is the case for small
amplitude motion.  A quantitative understanding of this effect is
still, however, lacking. Nonetheless, the general picture derived from
the linear theory, of an optimal compromise between the bending
required to break time reversibility and excessive damping suppressing
motion along the filament, remains valid. The most significant
difference we found was that there will come a point beyond which
increasing the wiggling frequency leads to a drop in efficiency.  The
theory, on the other hand, predicts that the efficiency remains
constant.

For realistic amplitudes of oscillation, we found that one-armed
swimming is, speed and efficiency-wise, a plausible strategy a
microorganism might use to get around. It is also a sight simpler than
the helical screw mechanism used by most bacteria. This requires a
rotary joint \cite{Bray}.
Nonetheless, while we stand open to correction, we have not been able
to identify a single organism that actually adopts this strategy.
Perhaps the most interesting question surrounding the one armed
swimmer is thus: why doesn't it exist?  Based on our results, we
suggest two hypotheses.  First, localized bending of the tail
requires implausibly high energy densities.  Second, the existence of
an evolutionary barrier. It is useless trying to swim with a short or
slow moving tail. Note that at small Sp (that is, low frequency and or
a short appendage) there is nothing to be gained in terms of motility.
This is not the case for either the helical screw mechanism, commonly
used by bacteria, or the traveling wave, used by spermatozoa. Both of
these give a {\em maximum} swimming speed and efficiency at low sperm
numbers.

Regarding experimental studies of in vitro motility, one main problem
so far is that the force involved has been too small to be detectable
with an optical trapping experiment. This limitation could be resolved
simply by time, as it is reasonable to expect that the resolution of
experiments will improve.  On the other hand, by means of the model
one could try to find the region of the parameter space where this
force is expected to be highest, and try to design an ``optimal
experiment'' where the motility could actually be quantified.

We would like to thank Catalin Tanase, Marileen Dogterom and Daan
Frenkel for discussion and help.

\section*{Appendix}
\subsection{Unequal friction coefficients as a condition
  for motility} It is possible to show that there can be no movement
if the viscous drag coefficients are equal.  Most conveniently, we work
with the the discrete model.  Since the discrete model produces the
continuum result in the limit that the number of beads $N$ goes to
infinity, there is no loss of generality (so long as the answer does
not depend on $N$).  The equation of motion for the center of mass is
\begin{equation}
   m \frac{ d \bar{{\bf v}} }{dt} = \sum_{i=1}^N {\bf f}_i
   \label{eq:cm}
\end{equation}
where
\begin{math}
  \bar{{\bf v}} = \frac{1}{N} \sum_{i=1}^N {\bf v}_i
\end{math}
is the center of mass velocity, \(m\) is the total mass.
The total force \({\bf f}_i \) on each bead consists of a bending
force \({\bf f}_{iB} \) a tension force \({\bf f}_{iT}\) an hydrodynamic
force \({\bf f}_{iH}\) and an ``external'' force \({\bf f}_{iX}\) which
accounts for the external drive. We know that by definition

\begin{displaymath}
  \sum_{i=1}^N {\bf f}_{iB} = \sum_{i=1}^N {\bf f}_{iT} = 0 
\end{displaymath}
Now, the external periodic force is applied only at one extremity, so that
\begin{math}
\sum_{i=1}^n {\bf f}_{iX} = {\bf f}_1(t)
\end{math}
and equation \ref{eq:cm} can be written as
\begin{displaymath}
  m \frac{ d \bar{{\bf v}} }{dt} = \sum_{i=1}^N {\bf f}_{iH} + {\bf f}_1(t)
\end{displaymath}
Integrating on a cycle we get
\begin{displaymath}
  m  \Delta \bar{\bf v}_{\tiny \textrm{cycle}} = \int_{ \tiny
  \textrm{cycle}} \textrm{dt } \sum_{i=1}^N {\bf f}_{iH} 
\end{displaymath}
The hydrodynamic force on bead \(i\) is written in the form
\begin{math}
  {\bf f}_{iH} = -\xi_{i \parallel}{\bf v}_{i \parallel} -\xi_{i
  \perp}{\bar{\bf v}}_{i \perp} 
\end{math}
(with \( \xi_{i \times} = \xi_{\times} / N \)). Thus, the effective drag on one
particle depends on the local configuration of the filament shape.

If the two friction coefficients
are the same \(\xi_{i \parallel} = \xi_{i \perp} = \xi_i\) then

\begin{equation}
   m  \Delta\bar{\bf v}_{\tiny \textrm{cycle}} = \xi  \int_{\tiny
   \textrm{cycle}} \textrm{dt } \bar{\bf v} 
   \label{eq:cycle}
\end{equation}

which necessarily leads to zero (or decaying to zero) global velocity. 
On the other hand, if the two drags are different, the right hand side
integral in equation \ref{eq:cycle} can be written as 
\begin{displaymath}
\int_{\tiny \textrm{cycle}} \textrm{dt } \xi_*(t) \bar{\bf v}(t)
\end{displaymath}  
where \(\xi_*\) is an effective drag which depends on time through the
configuration of the filament. This integral in general gives a number
once, in the spirit of resistive force theory, the configuration is
plugged in, and swimming is not, therefore, precluded.

\subsection{Analytical computation of the mean swimming
  velocity in the the small angular deviation approximation}

Her we outline the procedure adopted to
calculate analytically the average of the swimming speed using the
small deviation approximation. This calculation largely follows
the methodology used in~\cite{julicher} on a different model.

We can define the time average of the swimming speed (projected along
its only nonzero component along the \(  \hat{\mathbf{x}} \) direction) as

\begin{displaymath}
  \langle v \rangle = \lim_{t \rightarrow \infty} \frac{1}{t} \int_0^L
  \textrm{dt } \partial_t \mathbf{r} \cdot \hat{\mathbf{x}} 
\end{displaymath}

The expression for \( \partial_t \mathbf{r} \) can be obtained from
equation \ref{eq:motion} in terms of the local angle \( \psi \) as

\begin{displaymath}
  \partial_t \mathbf{r} = \frac{1}{\xi_{\perp}} \mathbf{n} (-k
  \partial_s^3 \psi + \tau \partial_s \psi) +
  \frac{1}{\xi_{\parallel}} \mathbf{t} (k \partial_s^2 \psi \partial_s
  \psi +\partial_s \tau)
\end{displaymath}

Fixing a reference frame one can consider the ``comoving'' frame with
respect to the filament, and expand \(\psi\) and \(\tau\), together with
the absolute displacements \(h\) and \(u\), and the swimming speed \(
\langle v \rangle\) as in formula \ref{eq:epsilon}. Following this
reasoning we can rewrite in vector notation the formula above for \(s
= 0\) as

\begin{equation}
  \begin{array}[h]{c}
    (\epsilon \langle v_1 \rangle + \epsilon^2  \langle v_2 \rangle, 0)
    + \partial_t ( \epsilon u_1(0) + \epsilon^2 u_2(0), 
    \epsilon h_1(0) + \epsilon^2 h_2(0)) = \\ \\
    =  \left[\frac{1}{\xi_{\perp}} \epsilon^2 k \psi_1 \partial_s^3
      \psi_1 + \frac{1}{\xi_{\parallel}} (k\epsilon^2 \partial_s^2 \psi_1
      \partial_s \psi_1 + \epsilon^2 \partial_s \tau_2) \right]_0
    \hat{\mathbf{x}} + \left[ \frac{1}{\xi_{\perp}} (-k\epsilon
      \partial_s^3 \psi_1) \right]_0   \hat{\mathbf{y}}
  \end{array}
  \label{eq:tau2}
\end{equation}

where we stop the expansion to second order in \( \epsilon
\). Expressing the equality for the different powers of \( \epsilon \)
one gets \(  \langle v_1 \rangle = 0 \) and
\begin{displaymath}
   \langle v_2 \rangle + \partial_t u_2(0) = - \psi_1(0) \partial_t
   h_1(0) + \frac{1}{\xi_{\parallel}} (k \partial_s \psi_1(0)
   \partial_s^2 \psi_1{0} + \partial_s \tau_2(0))  
\end{displaymath}

\(\tau_2(0)\) is obtained integrating equation \ref{eq:tau2}. Taking
into account the boundary conditions for \(h\), \(u\) and \(\tau\) one
gets to the expression
 
\begin{displaymath}
 \langle v_2 \rangle + \partial_t u_2(0) = \frac{\xi_{\perp} -
 \xi_{\parallel} }{\xi_{\parallel} L} \int_0^L \textrm{ds }\partial_s h_1
 \partial_t h_1 + \frac{1}{L} \int_0^L \textrm{ds } \int_0^s
 \textrm{ds'} \frac{1}{2} \partial_t (\partial_s h_1)^2 
\end{displaymath}

which, plugging in the analytical solution of (\ref{eq:wig}) and time
averaging, gives (\ref{eq:velocity}).

\end{document}